# Magnetic field dependent cycloidal rotation in pristine and Ge doped $CoCr_2O_4$

**N. Ortiz Hernandez[1*], S. Parchenko[1*] ⁑, J. R. L. Mardegan[1,4], M. Porer[1], E. Schierle[2], E. Weschke[2], M. Ramakrishnan[1], M. Radovic [1], J. A. Heuver[3], B. Noheda[3], N. Daffe[1], J. Dreiser[1], H. Ueda[1] and U. Staub[1**]**

*1 Swiss Light Source, Paul Scherrer Institute, Forschungsstrasse 111, 5232 Villigen PSI, Switzerland.*

*2 Helmholtz-Zentrum Berlin für Materialien und Energie GmbH, Albert-Einstein-Strasse 15, 12489 Berlin, Germany.*

*3 Zernike Institute for Advanced Materials, University of Groningen, Nijenborgh 4, 9747AG, The Netherlands.*

*4 Deutsches Elektronen-Synchrotron,* Notkestrasse 85, 22607 Hamburg, Germany.

** Authors with equal contribution.*

*⁑ Present address: Laboratory for Mesoscopic Systems, Department of Materials, ETH Zurich, 8093 Zurich, Switzerland*

***Corresponding author: urs.staub@psi.ch*

## Abstract

*We report a soft x-ray resonant magnetic scattering study of the spin configuration in multiferroic thin films of $Co_{0.975}Ge_{0.025}Cr_2O_4$ (Ge-CCO) and $CoCr_2O_4$ (CCO), under low- and high-magnetic fields, from 0.2 T up to 6.5 T. A characterization of Ge-CCO at a low magnetic field is performed and the results are compared to those of pure CCO. The ferrimagnetic phase transition temperature $T_C \approx 95$ K and the multiferroic transition temperature $T_S \approx 27$ K in Ge-CCO are comparable to those observed in CCO. In Ge-CCO, the ordering wave vector (qq0) observed below $T_S$ is slightly larger compared to that of CCO, and, unlike CCO, the diffraction intensity consists of two contributions that show a dissimilar x-ray polarization dependence. In Ge-CCO, the coercive field observed at low temperatures was larger than the one reported for CCO. In both compounds, an unexpected reversal of the spiral helicity and therefore the electric polarization was observed on simply magnetic field cooling. In addition, we find a change in the helicity as a function of momentum transfer in the magnetic diffraction peak of Ge-CCO, indicative of the presence of multiple magnetic spirals.*

## 1. Introduction.

Magnetoelectric (ME) multiferroics are of enormous interest from a technological perspective for designing new functionalities such as using electric fields to manipulate magnetic order [1]-[2]. Of special interest are type II multiferroics, where magnetic ordering drives the electric polarization with both order parameters being strongly coupled. This strong coupling enables switching the polarization by a magnetic field or the magnetization by an electric field, which is energetically more efficient.

$CoCr_2O_4$ (CCO) is one of the few known ME multiferroics of type II that exhibits a net magnetization due to its ferrimagnetic state [3]. CCO crystallizes in a spinel structure ($AB_2O_4$), having cubic symmetry ($Fd\overline{3}m$) with a lattice constant of 8.33 $\dot{A}$ [3] in bulk. The $Co^{2+}$ ions sit on the tetrahedral coordinated A sites and $Cr^{3+}$ on the octahedral coordinated B sites, subdivided into B1 and B2 sites. This material has been well characterized in both bulk [4]-[7] and thin film [8]-[12] forms. Three magnetic phases have been found below room temperatures for bulk. Below $T_C \approx 93$ K, where CCO becomes ferrimagnetically ordered [4]. In this phase, uncompensated magnetic sublattices of $Co^{2+}$ and $Cr^{3+}$ yield in a remanent net magnetization. Additionally, a spiral short-range order is reported to coexist within the long-range ferrimagnetic order [13]. In fact, the two magnetic sublattices B1 and B2 couple antiferromagnetically to each other with different opening angles for the cone, resulting in a net magnetic moment that is antiparallel to the sublattice A. This produces a net magnetization along [001] direction. Below $T_S \approx 26$ K, CCO gest an additional long-range magnetic spiral component, represented by an incommensurate modulation wave vector (*qq0*), with $q \approx 2/3$ reciprocal lattice units (r.l.u.). This transverse conical magnetic structure induces a ferroelectric polarization along the [$\overline{1}$10] direction [4]. Around $T_F \approx 15$ K, yet another magnetic phase transition to a commensurate spiral phase has been reported [4]-[7], the occurrence of which remains controversial. Choi [5] and Chang [6] reported the occurrence of a commensurate wave vector (2/3 2/3 0) and two additional incommensurate satellites in bulk CCO, with the new incommensurability being along the [110] and [1$\overline{1}$0]. A more recent work by Windsor [8] reported a single incommensurate spiral below $T_F$ in an epitaxially grown strained film. However, in this study, the width of the observed magnetic diffraction peak might have been too large to resolve additional long wavelength satellites.

The polarization direction in CCO is directly related to the helicity of the spin spiral as given by Katsura [14]: $\boldsymbol{P} \propto \hat{\boldsymbol{e}}_{ij} \times (\boldsymbol{S}_i \times \boldsymbol{S}_j)$, where $\boldsymbol{S}_{i,j}$ is the spin canting in the neighboring sites *i* and *j,* and $\hat{\boldsymbol{e}}_{ij}$ is the unit vector connecting the two sites which is parallel to the magnetic ordering wave vector *Q* = *(qq0).* For the case of thin films, the $\boldsymbol{Q}$ direction (out of the film surface) being fixed due to a slight tetragonal distortion caused by lattice mismatch with the substrate, the above relation implies that reversing the spin spiral results in a reversal of polarization and vice versa.

In this paper, we examine the effect of doping a small fraction of non-magnetic Ge on the long-range magnetic order in CCO. We compare the magnetic properties of epitaxial films of Ge doped CCO with pure CCO in the multiferroic phase. We also explore the behavior of the magnetic spin spiral in these systems under high magnetic fields. For this, we use resonant soft x-ray scattering (RSXS), an excellent technique to study complex magnetic structures. Recently, RSXS has been employed to investigate oxide materials, and particularly, multiferroics [15]-[21]. RSXS has the advantage of being element and orbital specific while probing long-range electronic ordering phenomena. Moreover, RSXS offers a high sensitivity in observing magnetic ordering schemes, even for small sample volumes [22]-[24].

## 2. Experiments.

Thin films of CCO and Ge-doped CCO were grown by pulsed laser deposition, monitored in-situ by reflection high-energy electron diffraction. The CCO thin films were grown with a thickness of ~80 nm on [110]-oriented MgO substrates. A more detailed description can be found in [11]. The same growth parameters were employed for the Ge doped CCO with 2.5% Ge doping, grown on [110]-oriented MgO substrates, resulting in $Co_{0.975}Ge_{0.025}Cr_2O_4$ (Ge-CCO) films with a thickness similar to CCO.

### 2.1. Resonant magnetic soft x-ray scattering (reflectivity and diffraction) under low magnetic fields.

X-ray magnetic circular dichroism (XMCD) in reflectivity mode and magnetic diffraction data have been collected at the RESOXS end station [25], at the X11MA beamline [26] of the Swiss Light Source (SLS). Intensities of reflected circularly polarized x-rays were collected at θ = 5° incidence for photon energies around the Co $L_{2,3}$ edges to obtain the XMCD signal in reflection mode (see experimental layout in Figure 1). For the magnetic diffraction experiment, *q*-scans were performed at 780 eV (Co $L_3$ edge). Both circular and linear x-ray polarizations were used. Data were collected with an IRD AXUV100 photodiode, covered by a 400 nm thick Al filter to suppress visible light and secondary electrons. The sample was field cooled (FC) in an external field of 0.2 T along the [001] direction from 300 K to 8 K prior to data collection.

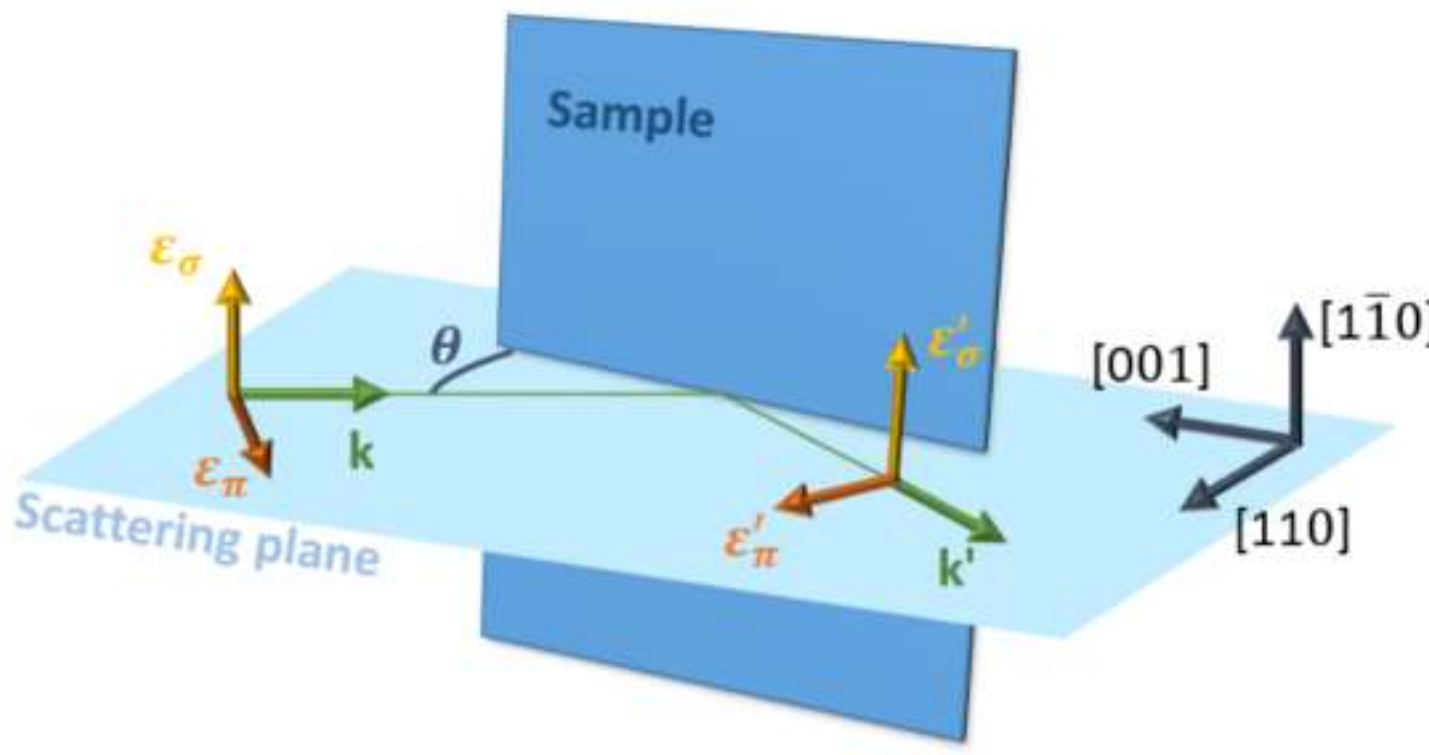

*Figure 1. Experimental geometry used in RESOXS end-station.*

### 2.2. XMCD on Ge-CCO collected by X-ray exited optical luminescence (XEOL).

XMCD and magnetic hysteresis measurements were carried out at the X-Treme beamline [27] of the Swiss Light Source using XEOL [8][28], taking the advantage of the insulating character of the samples and the luminescence of the substrates. For luminescent substrates, XEOL effectively measures the XAS in transmission mode [29]. Hysteresis loops were collected at the energy with the largest XMCD around the Co $L_3$ edge (777.5 eV) for various temperatures. An incident angle of 30° with respect to [001] direction was chosen. The sample was field cooled from room temperature to 10 K in a field of -0.2 T along [001].

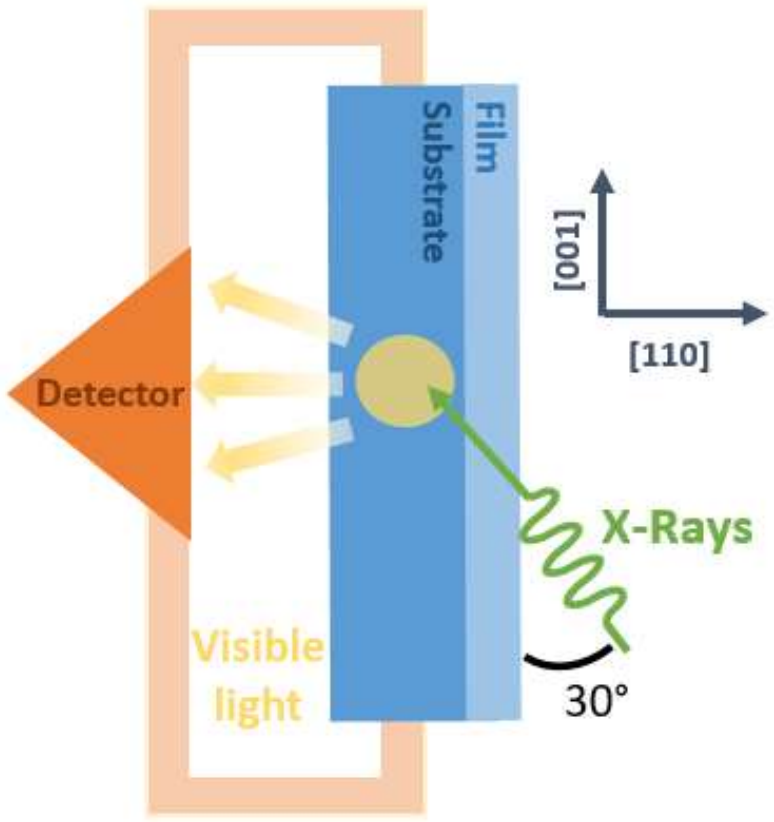

*Figure 2. Layout of XEOL experiment.*

## 2.3. Resonant soft x-ray scattering on Ge-CCO and pure CCO under high-magnetic fields.

Resonant magnetic x-ray diffraction measurements at the Co $L_3$ absorption edge have been carried out on the high-field diffractometer at the UE45_PGM1 beamline of BESSY II synchrotron, at Helmholtz Zentrum Berlin [30]. Incident linear and circular polarized x-rays were used at energies around the Co $L_{2,3}$ edge. A dome was built with 400 nm thick Al foil (same used in section 2.1), and placed above the sample to block visible light and secondary electrons. Both CCO and Ge-CCO samples were simultaneously mounted on the same holder to have the same experimental conditions. During field cooling, the magnetic field was applied along [001]. Since the magnets are not rotated during the scans, the magnetic field direction changes with respect to the surface during a scan.

Figure 3 displays a sketch of the experimental geometry. The magnetic field is created by four superconducting magnet coils, which are rotatable with respect to the sample. As the magnetic diffraction (*qq0*) peak requires a large scattering angle and the diffraction peak is very broad with a width of ≈ 45° in total scattering angle $2\theta$, the magnets were rotated by $\alpha$ = 32.5° with respect to the incident beam to have the best possible scattering geometry. When the magnetic field points towards the detector, secondary electrons are deflected by the field resulting in an artificial increase of the signal that disappears in the absence of the magnetic field. In order to reduce this effect, the sample holder was charged for a few seconds by applying an electric field along the [110] direction. To suppress the specular reflectivity background in the magnetic diffraction signal, the $\theta$ angle was displaced a quarter of a degree from the specular condition.

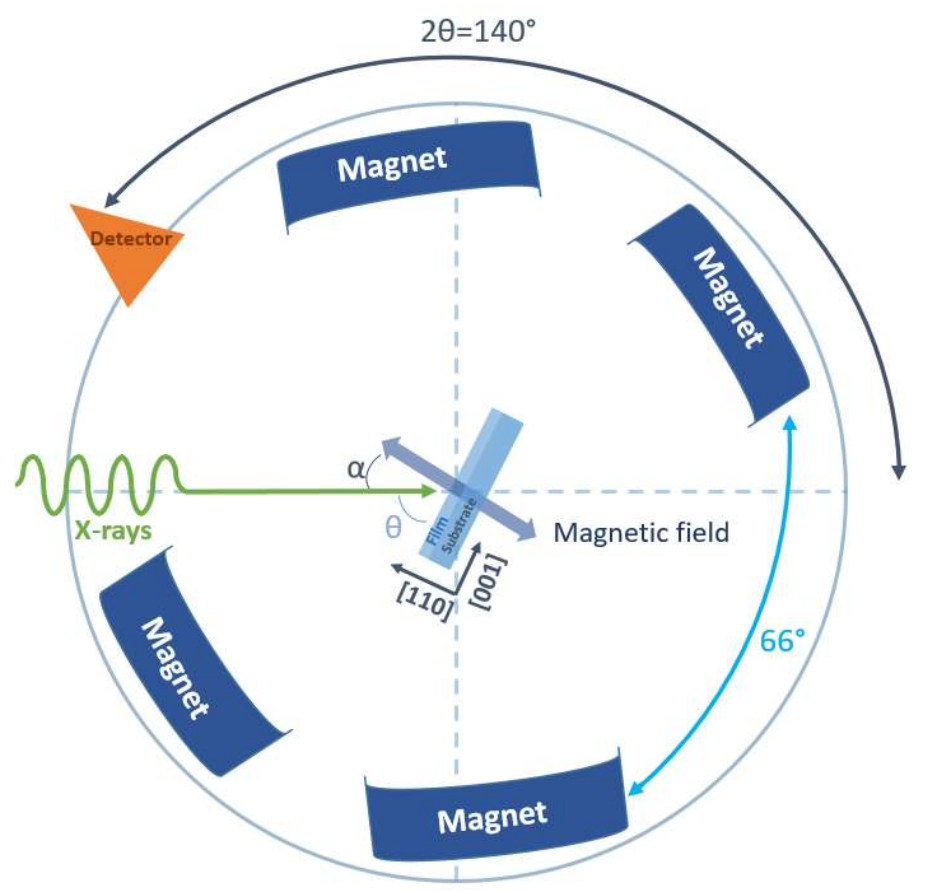

*Figure 3. Schematic representation of the experimental geometry. For (qq0) reflection, the detector was placed at 2θ = 140° with an incident angle (θ) of 69.75°. The magnetic field during measurements was rotated by an angle α = 32.5° from incoming X-ray beam.*

## 3. Results.

### 3.1. Resonant magnetic soft x-ray scattering (reflectivity and diffraction) under low magnetic fields.

Resonant x-ray scattering provides an element-specific measure of the electronic state and magnetic configuration of ions in a material [31]. In the case of a transition metal ion such as $Co^{2+}$, electric dipole excitations at $L_{2,3}$ absorption edges directly probe electron transitions from the 2*p* core to the 3*d* valance states. Thus, the spectra are sensitive to the electronic configuration of the 3*d* states and its spin configuration in the presence of a core hole [32]. The resonant magnetic scattering amplitude for a single ion can be expressed in the electric dipole approximation (E1-E1) as [32]-[33]

$$f_{E1E1} \propto [(\hat{\boldsymbol{\varepsilon}}'^{*} \cdot \hat{\boldsymbol{\varepsilon}})\boldsymbol{F}^{0} - i\hat{\boldsymbol{m}} \cdot (\hat{\boldsymbol{\varepsilon}}'^{*} \times \hat{\boldsymbol{\varepsilon}})\boldsymbol{F}^{1} + (\hat{\boldsymbol{\varepsilon}} \cdot \hat{\boldsymbol{m}})(\hat{\boldsymbol{\varepsilon}}'^{*} \cdot \hat{\boldsymbol{m}})\boldsymbol{F}^{2}] \quad (1)$$

where $\hat{\boldsymbol{\varepsilon}}$ and $\hat{\boldsymbol{\varepsilon}}'$ refer to the incoming and outgoing photon polarizations, $\hat{\boldsymbol{m}}$ is the unit vector of the magnetic moments and the terms $\mathbf{F}^{(n)}$ are scattering tensors of rank *n,* which depend strongly on energy. Scattered intensity is proportional to $|f_0 + f_{E1E1}|^2$, $f_0$ being the sum over the non-resonant amplitudes. For ferromagnetic order (e.g. solely the Co sublattice), only the second term in eq. (1) depends on the circular x-ray polarization and, to first order, is proportional to the magnetic moment. Hence, the circular dichroism can be approximated to be proportional to the magnetic moment. The circular dichroism is, therefore, large for small scattering angles when the sample is magnetized along the film plane and this is in the scattering plane, as it is in our case.

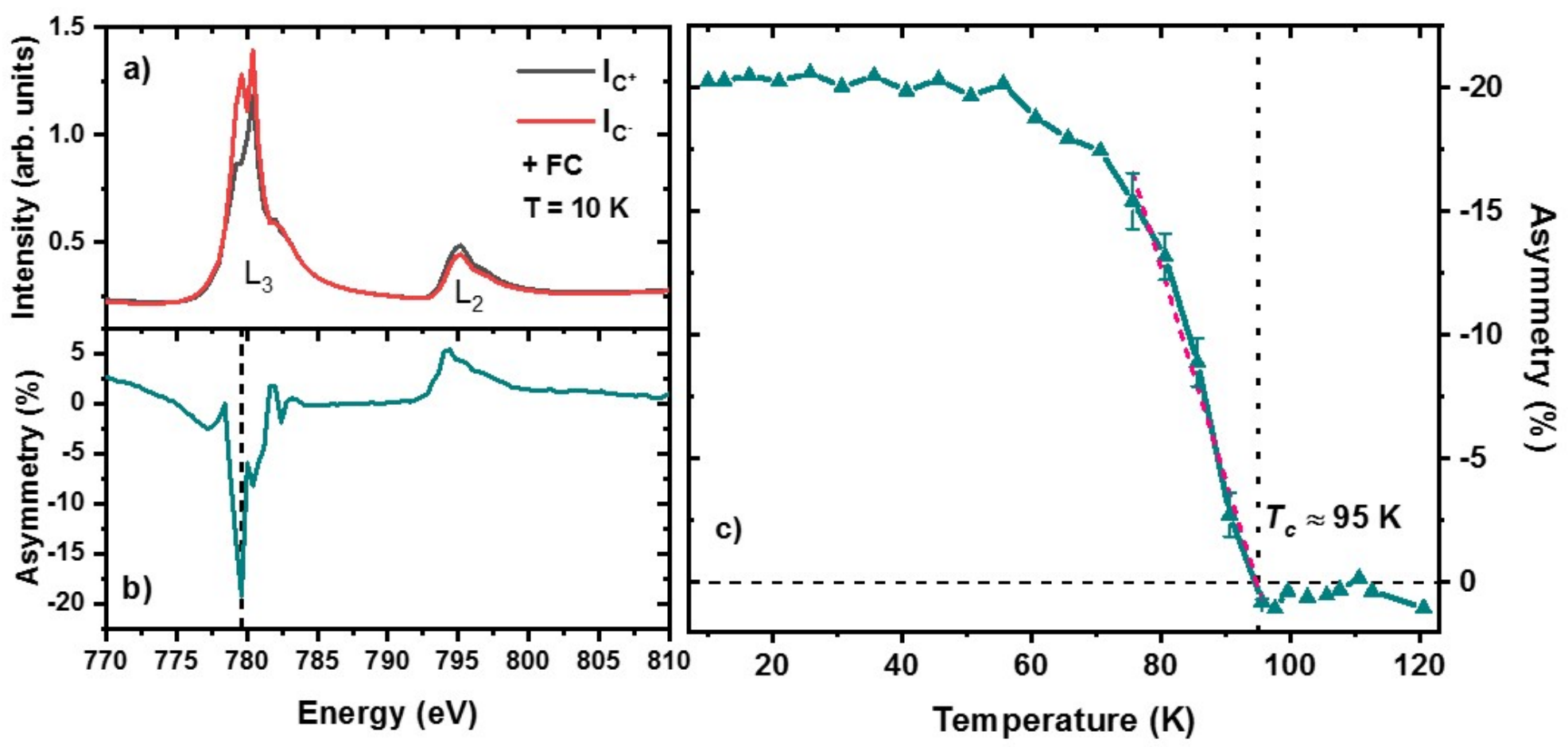

*Figure 4. (a) X-ray reflectivity of Ge-CCO at T = 10 K around the Co $L_{2,3}$ edges when field cooled at 0.2 T, for circular right (C+) and left (C-) polarizations. (b) Asymmetry as a function of energy. The dashed line indicates the energy with the largest asymmetry (in absolute values). (c) Magnetic asymmetry as a function of temperature measured at Co $L_3$ edge (energy of the dashed line in b), for a sample field cooled at 0.2 T. The pink dashed line estimates $T_C$. The small asymmetry offset above the Curie temperature is presumably due to an imperfect normalization procedure.*

Figure *4*a shows the energy spectra of the reflected beam at the Co $L_{2,3}$ edges for incident circular right ($C^+$) and circular left ($C^-$) polarized light for Ge-CCO. A clear contrast is observed between the two spectra. Figure *4*b shows the magnetic circular dichroism represented by the asymmetry defined as $A = \frac{I_{C^+} - I_{C^-}}{I_{C^+} + I_{C^-}}$. The dashed vertical line indicates the energy with maximal asymmetry, which is approximately 20% at 779.5 eV. Figure *4*c presents the asymmetry at 779.5 eV as a function of temperature. The black dashed lines indicate $T_C \approx 95$ K, which is estimated by a linear fit (indicated by the pink dashed line) and implies that it is a second order phase transition with a critical exponent $\beta \approx 1/2$. Above $T_C$ the circular dichroism disappears, indicating the transition to the paramagnetic phase, similar to that reported for pure CCO [8]. This suggests that a small amount of Ge doping does not affect the magnetic transition temperature of the ferrimagnetic phase significantly.

Figure *5*a displays the magnetic diffraction intensity of the (*qq0*) peak in Ge-CCO versus r.l.u. (bottom axis) and in total momentum transfer (*Q*) (top axis) for various temperatures. The peak maximum is, at least, 0.03 r.l.u. higher compared to pure CCO [8].

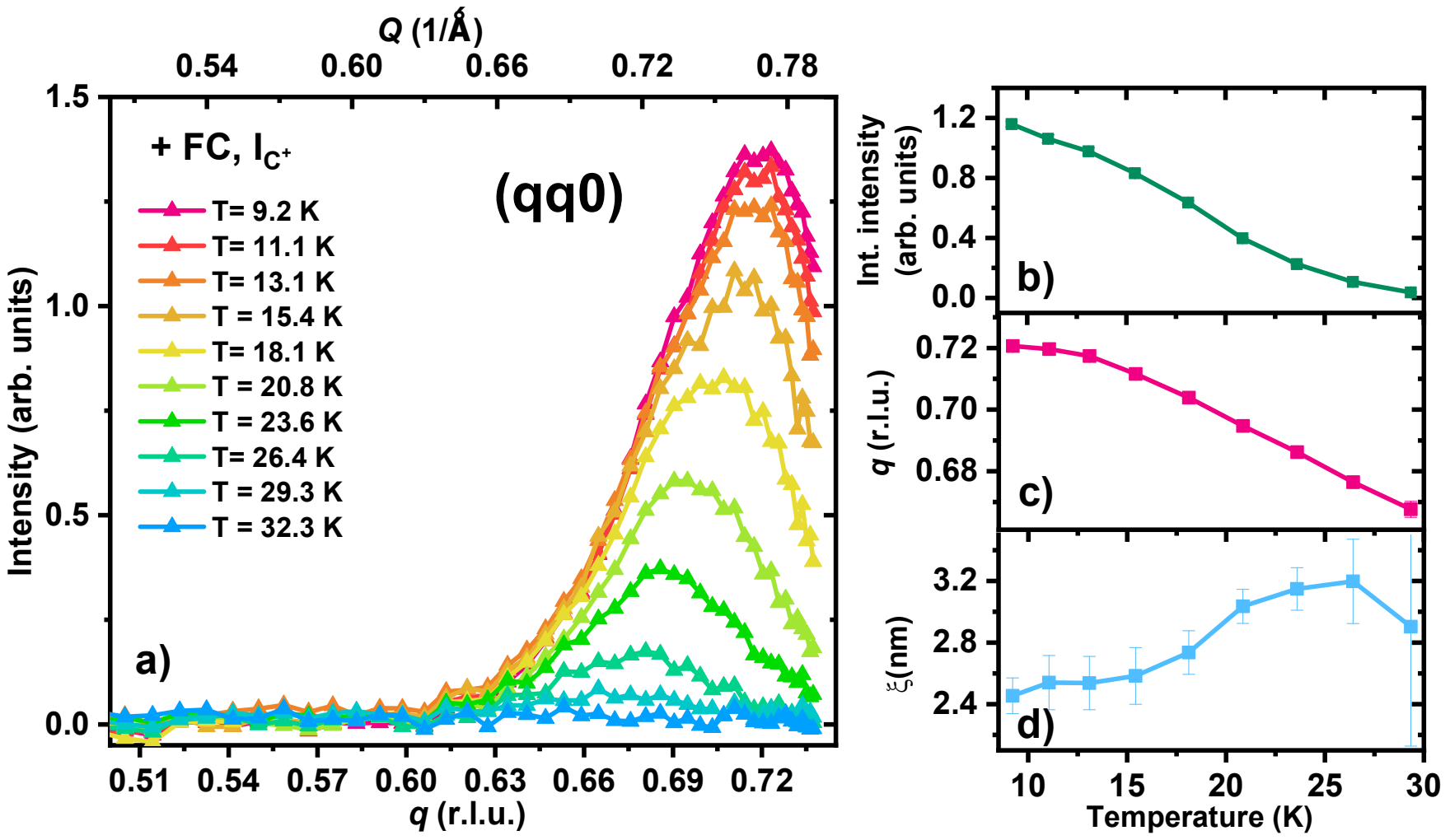

*Figure 5. (a) Temperature dependence of the magnetic diffraction peak (qq0) of Ge-CCO upon warming, after field cooling in 0.2 T and using C⁺ polarized x-rays at 780 eV. The data are presented as a function of both r.l.u. (bottom axis) and total momentum transfer (Q) (top axis). (b), (c) and (d) represent the temperature dependence of integrated intensity, modulation parameter q and the correlation length $\xi$ extracted from the data shown in (a).*

From a Gaussian fit in Figure *5*a, we extract the temperature dependence of the integrated intensity (Figure *5*b), the modulation parameter *q* (Figure *5*c) and the correlation length (Figure *5*d) calculated as $\xi = 2/FWHM$. Figure 5b shows that the static antiferromagnetic component appears around 27 K suggesting that the material is in the multiferroic state below this temperature, akin to pure CCO. Figure 5c shows an increase of the modulation parameter $q$ from 0.67 r.l.u. to 0.72 r.l.u. with reduction in temperature. We observe an increase of the correlation length with increasing temperature, which is in contrast to pure CCO [8]. This increase is prominent around 18 K, close to $T_F$ [4]-[7].

For comparison, we show in Figure 6, the magnetic diffraction (*qq*0) peak for circular and linear polarization with their corresponding circular ($I_{C+}$ - $I_{C-}$) and linear ($I_\pi$ - $I_\sigma$) dichroism for pure CCO (Figure 6 a- b) and Ge-CCO (Figure 6 c-d) at 779 eV collected at UE45_PGM1 beamline (BESSY II). For pure CCO, both linear and circular dichroism attain their maximum around 0.68 r.l.u. In contrast, Ge-CCO exhibits an extremum for linear dichroism around $q \approx 0.70$ r.l.u. (marked by the green dashed line), while the extremum for circular dichroism is around $q \approx 0.72$ r.l.u. (marked by a blue dashed line). This difference in $q$ for the extrema indicates that the diffraction peak is composed of more than a single magnetic contribution.

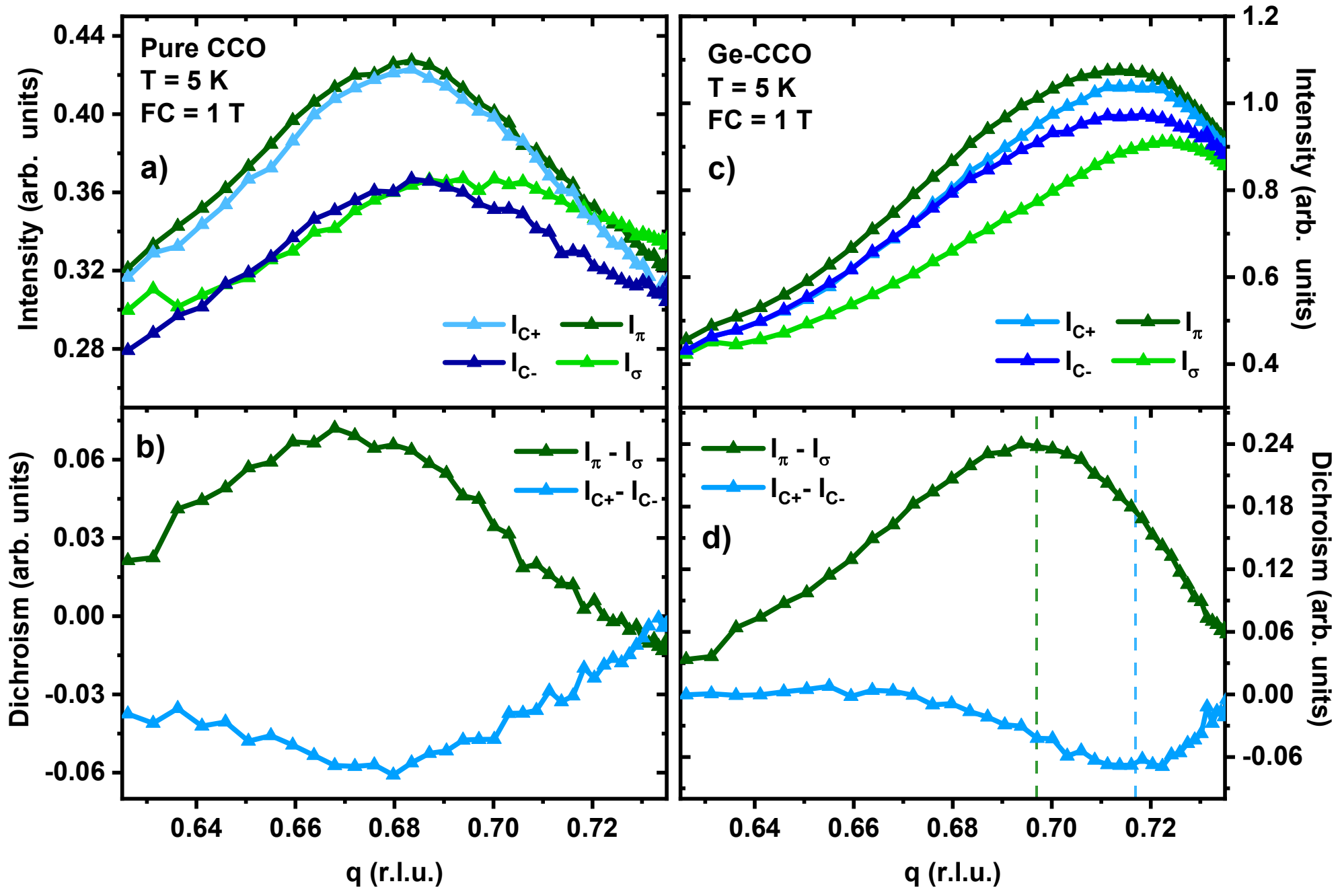

*Figure 6. Magnetic diffraction (qq0) peak for circular and linear polarizations with their respective linear and circular dichroism for CCO (a,b) and Ge-CCO (c,d). Data collected at 779 eV.*

### 3.2. XMCD on Ge-CCO collected by X-ray exited optical luminescence (XEOL).

The x-ray intensity transmitted through the film is measured by observing the XEOL signal and it can be described as:

$$I(z) = I_o \Lambda(E) e^{-\mu(E)z} \quad (2)$$

where $z$ is the thickness, $I_o$ is the incident intensity, $\mu$ is the energy dependent absorption coefficient of the sample, and $\Lambda(E)$ is the energy dependent efficiency function of XEOL for the substrate used (typical value for MgO at Co $L_{2,3}$ edge is 0.028 [29]). Taking into account the thickness of the substrate is ≈ 1 mm, we can assume that the entire signal is absorbed by the substrate, and therefore, the entire measured signal is XEOL [29] . Using equation 2, we calculate the XAS from the experimental XEOL data for each helicity of the x-ray ($\mu_\pm$), which are shown in Figure 7a. Figure 7b displays the corresponding XMCD ($\mu_+$-$\mu_-$) has a maximum at 777.5 eV.

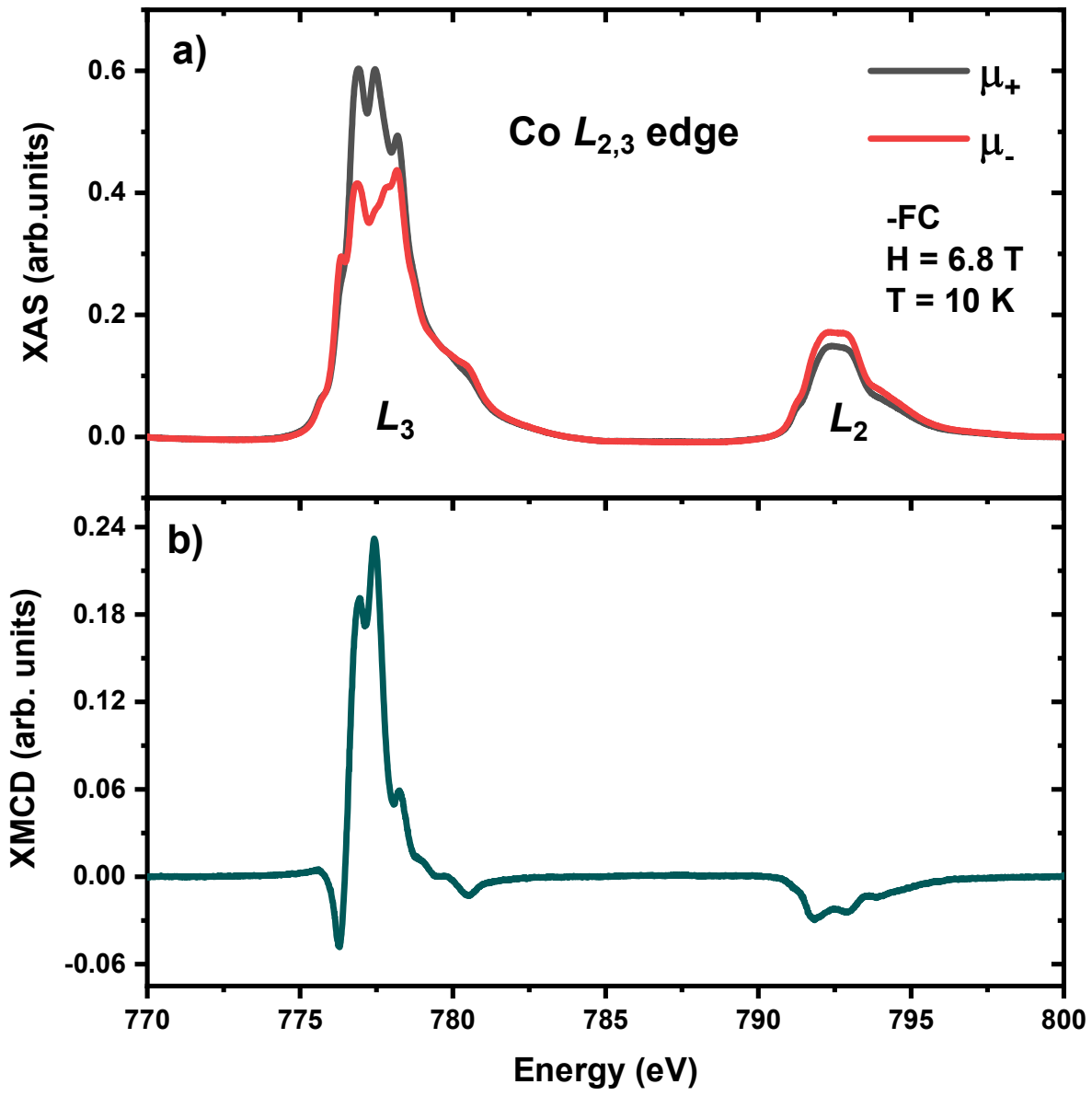

*Figure 7. (a) XAS of Co $L_{2,3}$ edge on Ge-CCO collected in XEOL mode at 10 K, under 6.8 T and FC -0.2 T. (b) XMCD data obtained from (a).*

Furthermore, hysteresis loops were recorded by means of XMCD at 777.5 eV for various temperatures, as shown in Figure 8. For each applied magnetic field, measurements below the absorption edge (770 eV) have been used for baseline correction. These data show that for $T \leq 40$ K, a magnetic field of 6.8

T is insufficient to saturate the magnetization, indicative of a very large coercivity and larger saturation field when compared to pure CCO thin film [8].

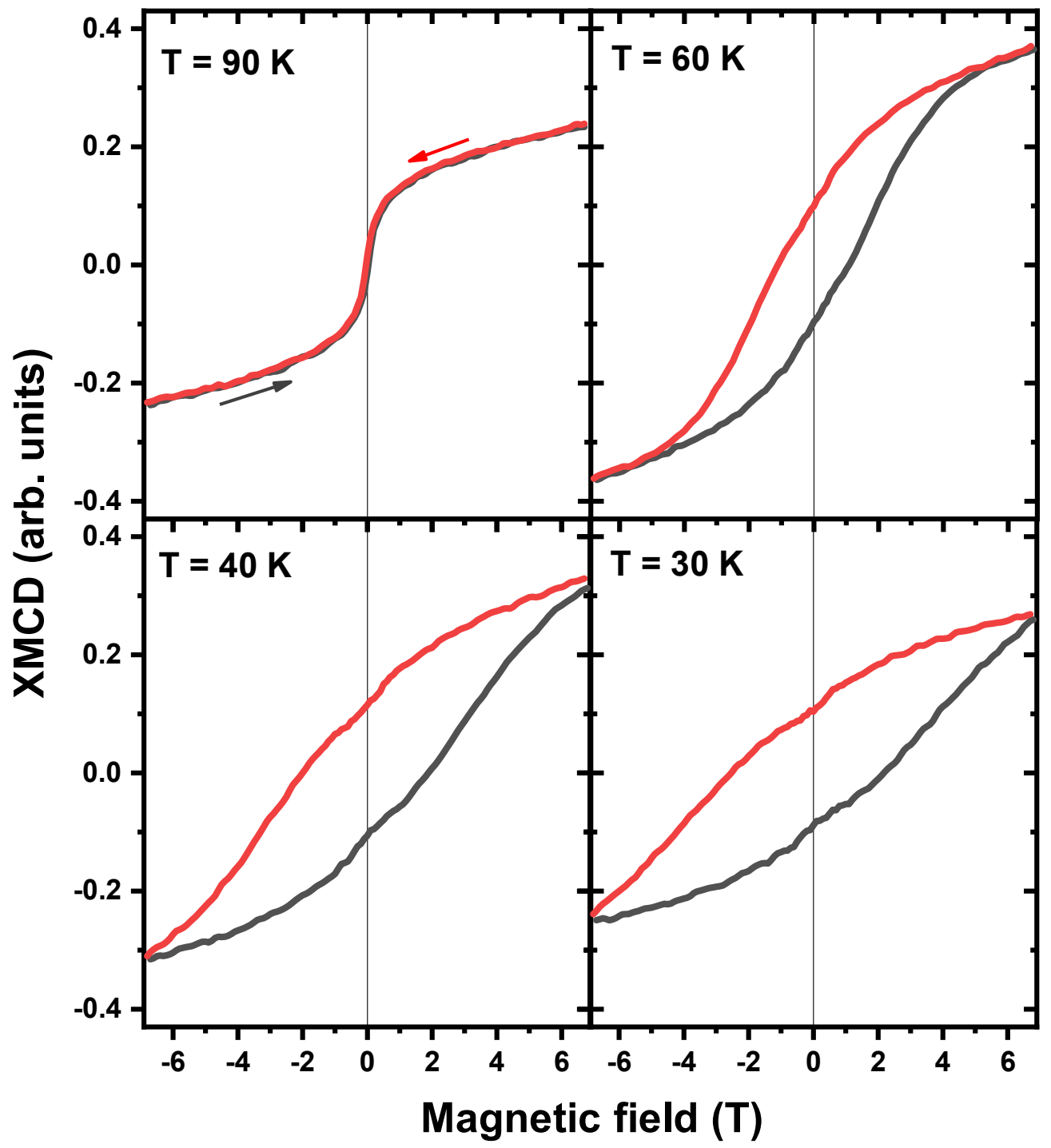

*Figure 8. Magnetic hysteresis loop of Co sublattice in Ge-CCO for various temperatures measured in XEOL mode.*

### 3.3. Resonant soft x-ray scattering on Ge-CCO and pure CCO under high-magnetic fields.

Figure 9 presents the energy dependence of the magnetic diffraction ($qq0$) peak ($q \approx 0.69$) on pure CCO and Ge-CCO for opposite circular polarizations. Both samples have been field cooled at -6 T from room temperature to 5 K and a magnetic field of -6 T was applied during the measurements.

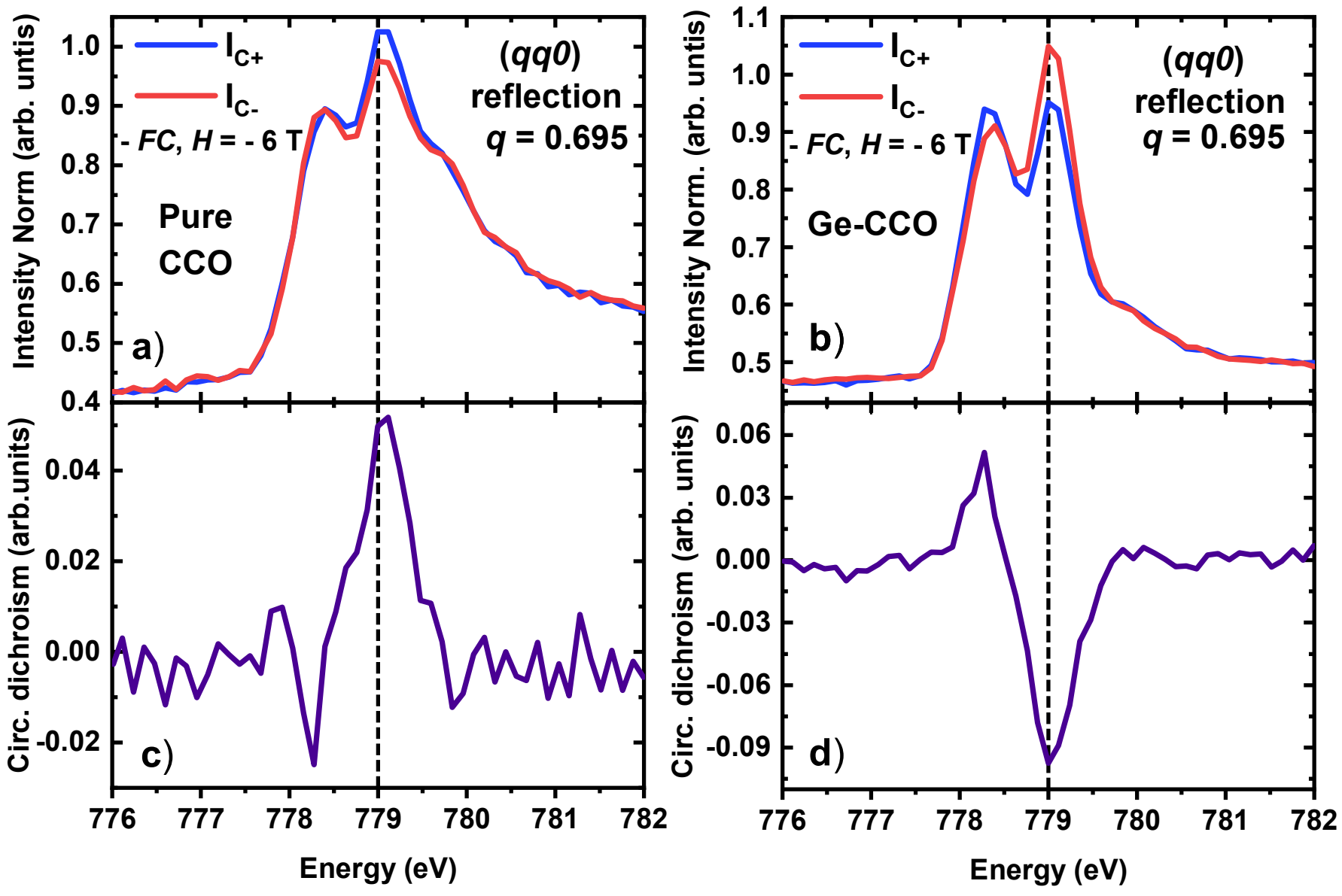

*Figure 9. Energy dependence of the magnetic diffraction* (qq*0) peak for* $C^+$ *and* $C^-$ *polarizations for (a) pure CCO and (b) Ge-CCO at 5 K, under applied of -6 T after FC at -6 T. Energy dependence of the circular dichroism of the* (qq*0) peak for (c) pure CCO and (d) Ge-CCO. Dashed lines indicate the energy corresponding to the extrema of circular dichroism.*

The energy dependence of the (*qq0*) reflection at *q* = 0.695 r.l.u. has similar features for both materials, with a maximum at 779 eV and a shoulder around 778.3 eV. However, in the case of pure CCO, the spectrum of the diffraction peak is broader and less pronounced than in Ge-CCO. The observed circular dichroism in Ge-CCO and pure CCO have opposite signs at 779 eV (Figures 9c and 9d).

To learn more about the observed circular dichroism, we studied the field dependence of the (*qq0*) peak at 779 eV. We collected two data sets with opposite applied fields for each sample. In the first data set, both materials were cooled under a field of -1 T from 120 K to 5 K and data were acquired while increasing the applied field in several steps from 0 T up to -6.5 T. For the second data set, the samples were cooled at 6.5 T from 60 K (which is fully sufficient for reversing the magnetization) to 5 K and same procedure was carried out, with the exception that the measurements were done while decreasing the applied field in steps from 6.5 to 0 T.

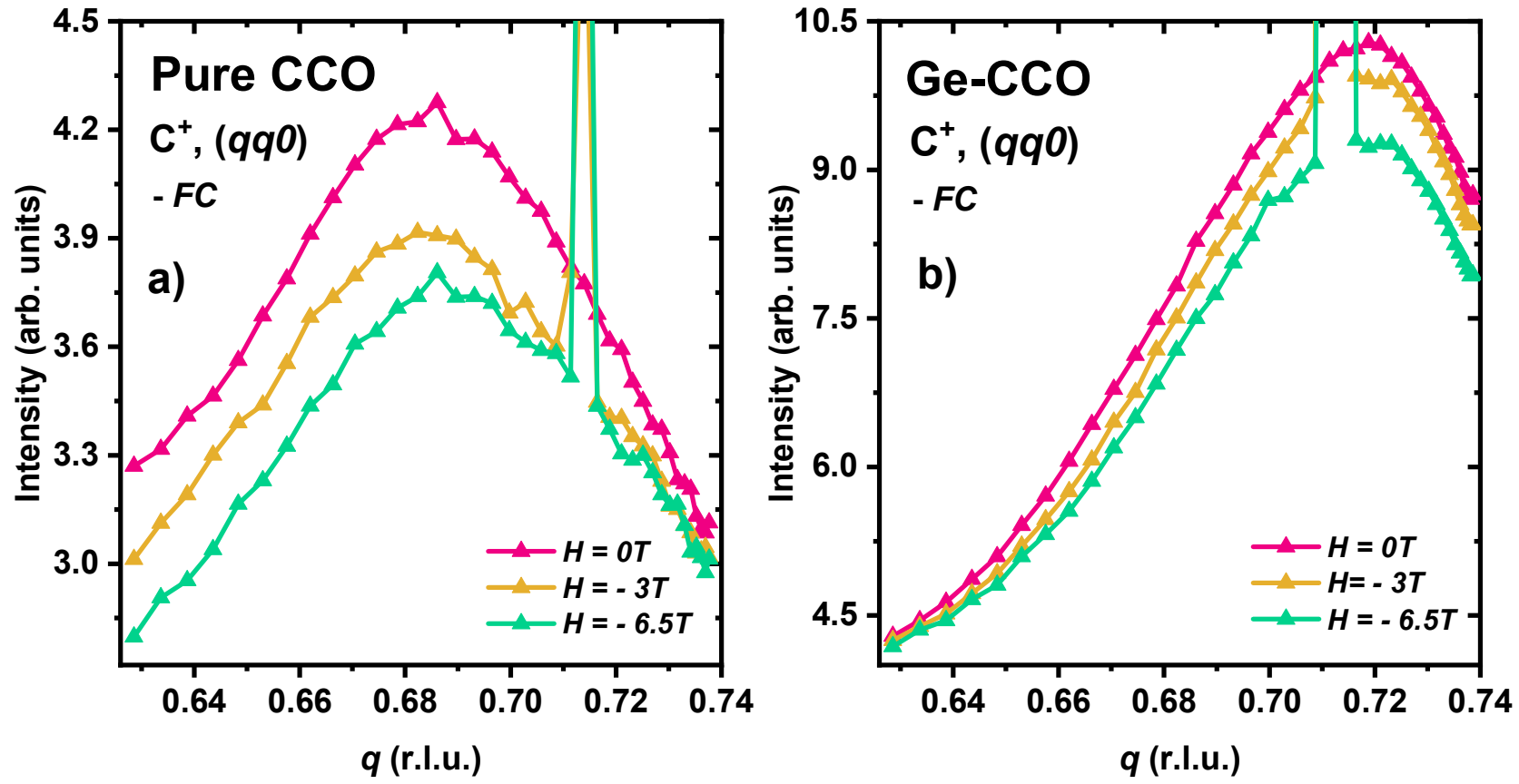

*Figure 10. Intensity of magnetic diffraction peak (qq0) of pure CCO (a) and Ge-CCO (b) for various magnetic fields at 779 eV. At 5 K, data collected with $C_+$ polarized light, in field cooled at -1 T. The origin of the artifact around q ≈ 0.712 is explained in the text.*

Figure 10 displays (*qq0*) peak under various magnetic fields on pure CCO (Figure 10a) and on Ge-CCO (Figure 10b) for the case of negative field cooling and incident C+ polarization. The sharp peak observed around *q* = 0.71 r.l.u. is an artifact caused by secondary electrons when the magnetic field points to the detector, as explained previously. The application of the field results mainly in a reduction and/or distortion of the (*qq0*) reflection intensity for both materials.

The circular dichroism observed in diffraction peak of a spin spiral defines directly the sign of the cycloidal rotation of the magnetic moments similar to those found in $TbMnO_3$ [34] or $DyMnO_3$ [35]. From now on, we refer to the circular dichroism of the magnetic diffraction (*qq0*) peak as helicity contrast, defined as $\frac{I_{C+} - I_{C-}}{\int (I_{C+}+I_{C-})dq}$.

In Figure *11*, we present the helicity contrast of the (qq0) peak for the different sets of measurements: for pure CCO with negative applied magnetic field (*Figure 11*a) and positive applied magnetic field (*Figure 11*b). The case of Ge-CCO is shown in *Figure 11*c and *Figure 11*d for negative and positive applied magnetic fields, respectively.

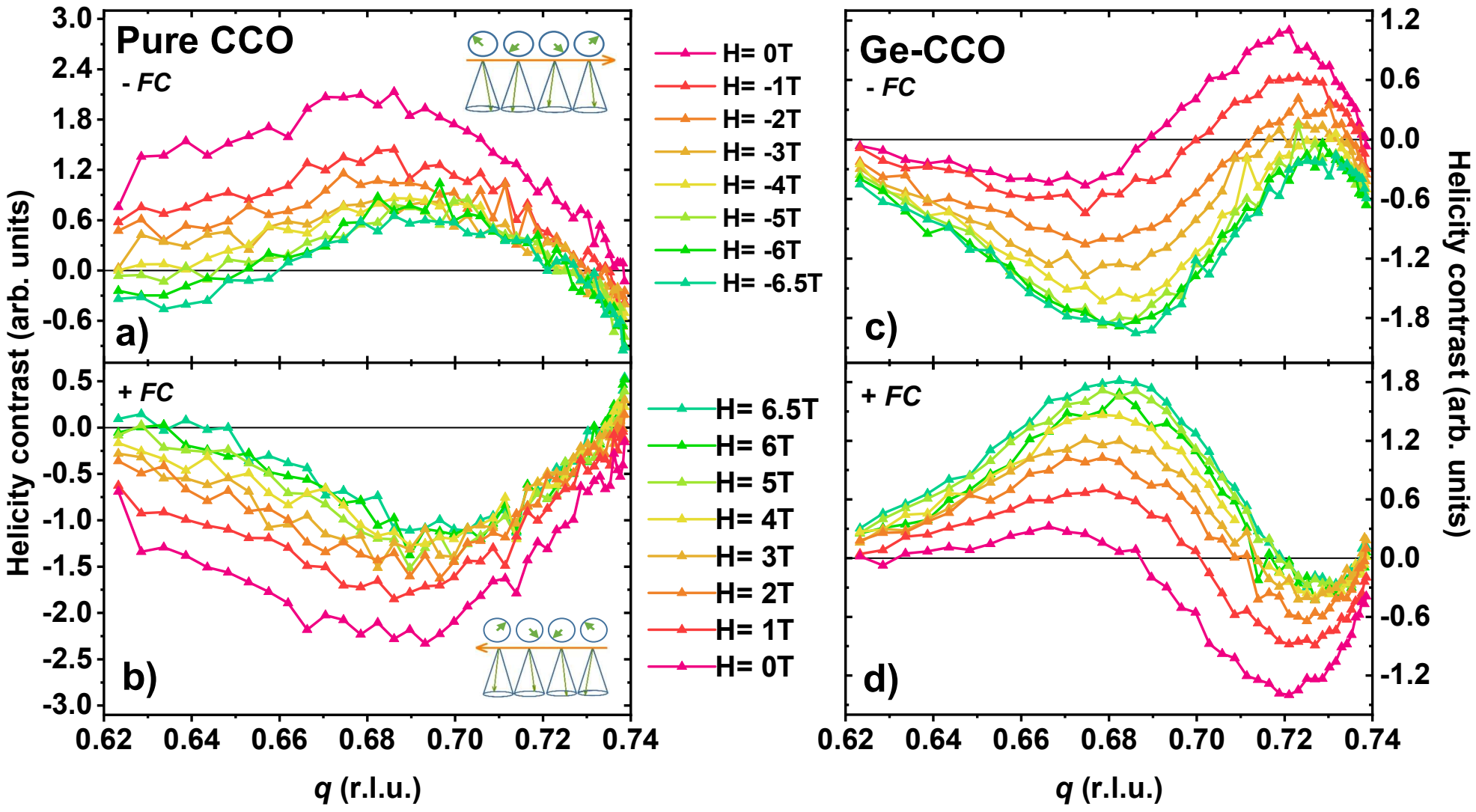

*Figure 11. Helicity contrast of the (qq0) peak for various magnetic fields at 5 K on (a,b) pure CCO and on (c,d) Ge-CCO at 779 eV. (a) and (c) panels are for negative magnetic field cases and (b) and (d) for positive magnetic field cases. Insets in (a,b) panels are the sketch of the magnetic modulation vector ( orange arrows) and magnetic moments (green arrows).*

While being significantly different in shape, the helicity contrast shows a mirror effect between opposite applied field directions for both materials (Figure *11*c-d). Therefore, both materials exhibit a direct correlation between the direction of magnetic field cooling and cycloidal rotation. Insets in Figure *11*a-b show the helicity sense, indicated by the orange arrows. In addition, the helicity contrast decreases for increasing fields for CCO, possibly due to a field dependent elliptical distortion of the cycloid. The behavior of Ge-CCO though is more complex. There are two extrema in the helicity contrast as a function of $q$. At $q_1 \approx 0.72$ r.l.u., we observe a maximum/minimum for positive/negative applied fields, similar to pure CCO. The helicity contrast is reduced when the magnetic field increases, irrespective of the magnetic field direction. At a lower $q$, the extremum around $q_2 \approx 0.68$ r.l.u. is enhanced at larger applied fields. The two extrema $q_1$ and $q_2$ show opposite helicity to each other. The sign change in the helicity contrast as a function of $q$ exhibited by Ge-CCO is not observed in pure CCO, which only possesses a single maximum or minimum depending on the sign of the applied field. The observation of two extrema in Ge-CCO may be interpreted as the appearance of a second cycloidal component with slightly smaller $q$ and opposite helicity. In order to extract field induced changes, the signal collected under $H$ = 0 T has been subtracted from the data collected under field, as shown in Figure *12*. Pure CCO shows a subtle extremum around 0.66 r.l.u., which reflects a small change in the peak shape. In Ge-CCO, however, a single clear maximum around $q$ = 0.69 r.l.u. is observed, indicating that the magnetic contribution at $q_1$, observed in the zero-field contrast is independent of the applied magnetic field.

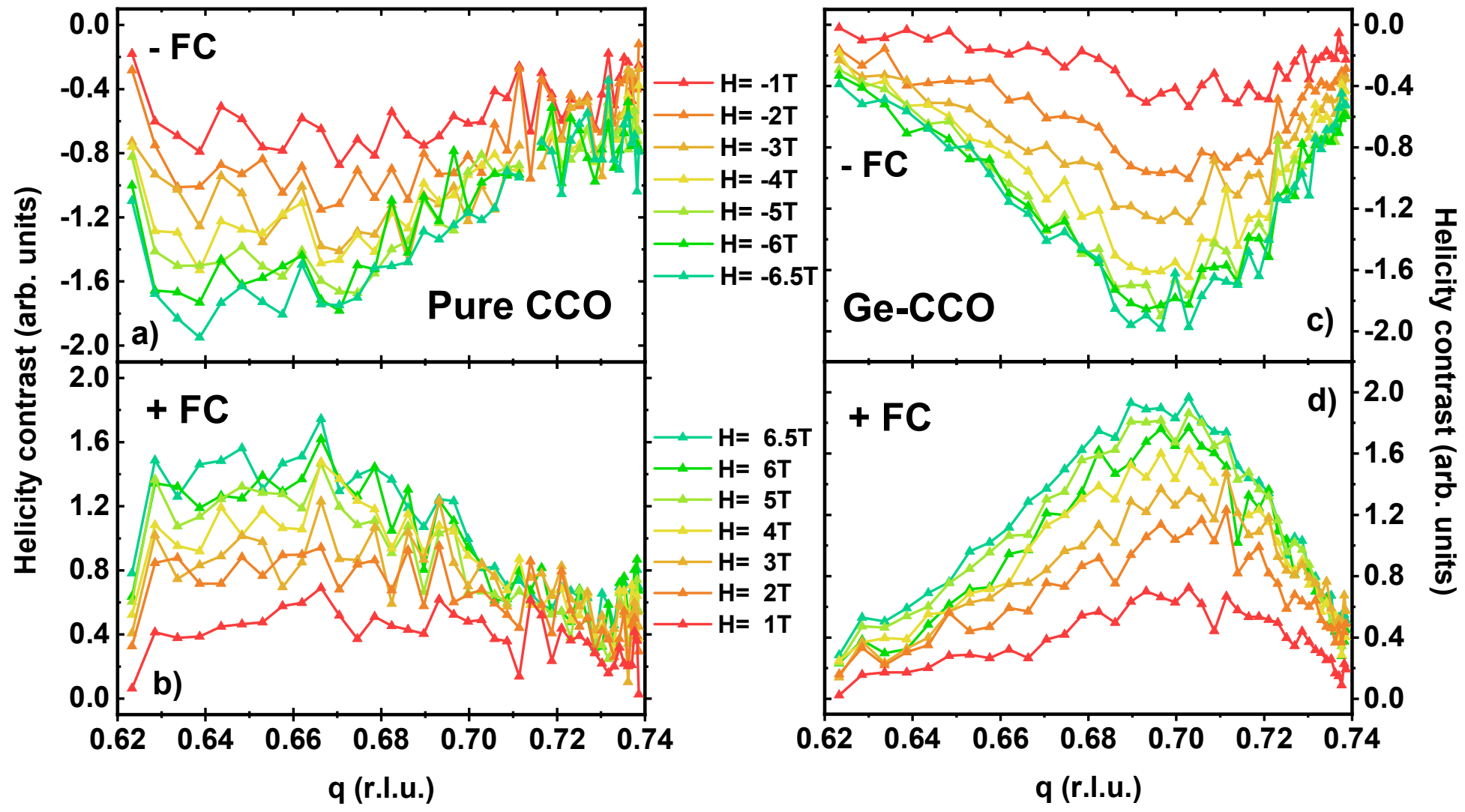

*Figure 12. Helicity contrast of the (qq0) reflection for various magnetic fields after subtracting the zero-field signal. Results of CCO (a,b) and Ge-CCO (c,d) for negative FC and positive FC, respectively.*

## 4. Discussion and conclusions.

The Ge doped CCO thin film exhibits magnetic transitions at temperatures similar to those observed in pure CCO [8], with $T_C \approx$ 95 K and the multiferroic phase appearing around $T_S \approx$27 K. However, the small fraction of Ge doping causes an increase of coercive and saturation field in Ge-CCO. Studying the (*qq0*) magnetic diffraction peak of Ge-CCO, we observe that $q$ is temperature dependent and goes from a commensurate value of 0.67 r.l.u. values at 27 K to incommensurate 0.72 r.l.u. at 9 K (Figure 5). These values are larger, at least by 0.03 r.l.u., than the ones reported for the CCO film [8]. Furthermore, an increase in the correlation length is observed with increasing temperature. This is possibly an indication that Ge-CCO undergoes a magnetic phase transition ($T_F$) around 18 K, similar to what has been reported in bulk CCO [4]-[7] but not observed in pure CCO films. This view is supported by the linear and circular dichroism extrema observed around $q \approx 0.70$ and $q \approx 0.72$ r.l.u. in Ge-CCO (Figure 6).These observations could be an indication of the coexistence of multiple spin cycloidal textures for temperatures below $T_F$ as found in bulk CCO [5]-[6], where the (*qq0*) magnetic peak splits into three different components below $T_F$.

The multiple spin cycloidal scenario in Ge-CCO gets further support from the results observed in high-magnetic field, where distinct pattern is observed in the helicity contrast possessing a maximum and a minimum at different values of $q$ (Figure *11*). As the diffraction contrast between circular polarizations relates directly to the sense of the cycloidal rotation, it indicates that the two extrema with opposite signs, at $q_1$ and $q_2$, represent two different modulation vectors close to (qq0) with opposite handed spin rotations. Interestingly, in CCO, the helicity contrast features just one maximum, whose intensity is reduced when magnetic field increases. This may simply reflect a reduction and/or a distortion (ellipticity) of the cone aperture angle (which defines the moment contribution of the cycloid) as the moments try to align with the magnetic field. For the case of Ge-CCO, $q_1$ behaves as CCO, while $q_2$ gets enhanced with increase in magnetic field. These results show that Ge-CCO has at least two types of spin cycloids oriented differently with respect to each other, or having a different length of the propagation vector with different helicity. Assuming the low temperature phase, below $T_F$, is characterized by the appearance of the commensurate (2/3 2/3 0) modulation and some satellite reflections, as reported in [5]-[6], the satellite at $q_1$ may be the order parameter of the low temperature phase for Ge-CCO films. The difference in behavior of Ge-CCO and CCO may be due to the fact that the CCO thin film has much lower $T_F < 5$ K because of the strain, which makes such a transition observable in bulk but not in the pure CCO films. The difference in $T_F$ between Ge-CCO and CCO implies that Ge doping, directly affects the tiny balance between the magnetic exchange interactions, while having negligible effect on the temperature of the ferrimagnetic and multiferroic ordering that are more sensitive of the overall scale of the magnetic exchange constants.

An interesting observation in our study is a direct correlation between the magnetic field cooling and the spin cycloidal rotation, shown in Figure *11*. As explained earlier, the circular dichroism of a magnetic diffraction signal directly relates to the helicity of the cycloidal rotation. According to the relationship of the electric polarization to the spiral rotation, $\boldsymbol{P} \propto \boldsymbol{e_{ij}} \times (\boldsymbol{S_i} \times \boldsymbol{S_j})$ [14], for a thin film, the reversal of the spin cycloidal leads to the reversal of the polarization. Here, we report the reversal of the spin cycloid and, as a consequence, the reversal of the polarization through only magnetic field cooling. Usually, in multiferroics, a combination of electric and magnetic fields is needed to achieve a single domain state. For our samples, it is unclear why magnetic field cooling alone produces a single multiferroic domain state. This may be due to the fact that the polarization direction in these thin films is well defined without the need for fixing it through an electric field cooling process. In general, we can propose a few mechanisms, which leads to a well-defined polarization direction. (1) A bias created by the difference of voltage applied prior to the measurements to charge the sample could define the polarization direction. This scenario is, however, unlikely since the applied voltage produces an electric field perpendicular to the sample surface (the [110] direction) which does not affect the polarization that lies along $[\bar{1}10]$. (2) X-rays polarize the sample, as reported by Schierle [35], leading to a defined cycloidal rotation domain. This scenario is again not applicable in our case since we observe a reversal of the cycloidal rotation, i.e. reversal in polarization, for opposite magnetic field cooling from above $T_c$. (3) Bias produced by inversion symmetry breaking due to strain at the interface is another possibility. However, the strain does not break inversion symmetry along the in-plane $[\bar{1}10]$ direction. (4) A plausible explanation could be a bias produced along $[\bar{1}10]$ by antisymmetric exchange, i.e. Dzyaloshinskii-Moriya interaction (DMI). DMI can produce weak ferromagnetism caused by canting of all the collinear AFM moments towards one direction, produced by spatially alternating DM vectors. In this case, inverting the external field may cause the inversion of all DM vectors. In any case, further investigations on the origin this effect is required to confirm such a hypothesis.

In summary, our x-ray investigation finds that Ge doping in CCO does not alter the main magnetic properties in the ferrimagnetic state nor the onset of the multiferroic phase. Despite the similarity in the temperature of phase transitions, the ground state of the Ge-doped film shows a more complex magnetic behavior below $T_S$ compared to the pure CCO films. We find the occurrence of a second cycloidal component in the magnetic structure, which is close to commensurate, which might represent the phase below $T_F$ = 15 K occurring in bulk CCO. Only one of the cycloids observed in the doped system is magnetic field dependent, although surprisingly, both reverse their helicity, which also represents an inversion of the electric polarization, for opposite field cooling.

## Acknowledgement.

We gratefully thank the X11MA and X07MA beamline staff for experimental support. The financial support of the Swiss National Science Foundation, and its National Center of Competence in Research, Molecular Ultrafast Science and Technology (NCCR MUST) No. 51NF40-183615 is acknowledged and M.R. and N. O.H. acknowledge financial support of the Swiss National Science Foundation (SNSF) (Sinergia project ’Toroidal moments’ No. CRSII2_147606 and No. 200021_169017, respectively). H.U. acknowledges financial support from the European Union’s Horizon 2020 research and innovation programme under the Marie Skłodowska-Curie grant agreement No. 801459 - FP-RESOMUS - and the Swiss National Science Foundation through the NCCR MUST. The research leading to this result has been supported by the project CALIPSOplus under the GRANT Agreement 730872 from the EU Framekwork Programme for Research and Innovation Horizon 2020.

## References.